\documentclass[12pt]{article}
\RequirePackage[OT1]{fontenc}

\usepackage{amsmath,amssymb,amsthm}
\usepackage{natbib}
  \bibpunct{(}{)}{;}{a}{,}{,}
\usepackage{graphicx}
\usepackage{bm}
\usepackage[vlined]{algorithm2e}
\usepackage[pdftex,colorlinks=true,linkcolor=blue,citecolor=blue,urlcolor=blue,bookmarks=false,pdfpagemode=None]{hyperref}
\usepackage{verbatim}
\usepackage[top=1.1in, bottom=1.1in, left=1in, right=1in]{geometry}
\usepackage{setspace}

\usepackage{fancyhdr}
\newcommand{\obs}{{\rm obs}}

\newcommand{\floor}[1]{\left \lfloor #1 \right \rfloor}

\newcommand{\Lihat}{\hat{T}_i^{\rm LaPRET}}
\newcommand{\Li}{T_i^{\rm LaPRET}}
\newcommand{\Oi}{T_i^{\rm obs}}
\newcommand{\toi}{T_i^{\rm event}}
\newcommand{\Ti}{T_i^{\rm treat}}
\newcommand{\kg}{{\rm kg}}
\newcommand{\m}{{\rm m}}
\DeclareMathOperator*{\argmax}{arg\,max}%

\graphicspath{{.}}
\begin{document}

\pagestyle{empty}
\title{Observational studies with unknown time of treatment}
\author{Guillaume W. Basse$^\dagger$ \and Alexander Volfovsky$^\dagger$ 
\and Edoardo M. Airoldi\thanks{Guillaume W.\ Basse is a graduate student in the Department of Statistics at Harvard University (gbasse@fas.harvard.edu). Alexander Volfovsky is an National Science Foundation Postdoctoral Fellow in the Department of Statistics at Harvard University (volfovsky@fas.harvard.edu). 
Edoardo M.\ Airoldi is an Associate Professor of Statistics at Harvard University (airoldi@fas.harvard.edu). 
This work was partially supported 
 by the National Science Foundation under grants 
  CAREER IIS-1149662, IIS-1409177, and DMS-1402235,
 and by the Office of Naval Research under grant 
  YIP N00014-14-1-0485. 
 Edoardo M.~Airoldi is an Alfred P. Sloan Research Fellow, and a Shutzer Fellow at the Radcliffe Institute for Advanced Studies.
The authors wish to thank Michael Els, Robert Haslinger, Mark Lowe, Sean Murphy, and Donald B.\ Rubin for providing data, comments, and valuable insights. 
~ $^\dagger\,$These authors contributed equally to this work.}}
\date{}

\maketitle
\thispagestyle{empty}

\newpage

\begin{abstract}
Time plays a fundamental role in causal analyses, where the goal is to quantify the effect of a specific treatment on future outcomes.
In a randomized experiment, times of treatment, and when outcomes are observed, are typically well defined.
In an observational study, treatment time marks the point from which pre-treatment variables must be regarded as outcomes, and it is often straightforward to establish.
Motivated by a natural experiment in online marketing, we consider a situation where useful conceptualizations of the experiment behind an observational study of interest lead to uncertainty in the determination of times at which individual treatments take place. 
Of interest is the causal effect of heavy snowfall in several parts of the country on daily measures of online searches for batteries, and then purchases. The data available give information on actual snowfall, whereas the natural treatment is the anticipation of heavy snowfall, which is not observed. 
In this article, we introduce formal assumptions and inference methodology centered around a novel notion of plausible time of treatment. These methods allow us to explicitly bound the last plausible time of treatment in observational studies with unknown times of treatment, and ultimately  yield valid causal estimates in such situations.\newline
\vfill

\end{abstract}

\newpage
\onehalfspacing
\tableofcontents
\doublespacing


\newpage

\pagestyle{fancy}
\setcounter{page}{1}



\section{Introduction}

Observational studies are common across many disciplines, 
including the health and social sciences where practitioners
are interested in assessing the causal effect of a non-randomized
treatment \citep{Rosenbaum:2002aa,Rosenbaum:2010aa}. For example, a drug company might solicit 
information on
the blood pressure of individuals following the 
voluntary ingestion
of a particular drug, or a fast food restaurant might record
observations on sales over time. The former might be interested
in the effect of the drug on blood pressure, while the latter
might be interested in the effect of the forecast of an adverse weather event on sales. 
In both cases, the data are gathered from historical records, thus without the ability
to randomize treatment. Because of the lack of randomization, 
naive estimators of the average treatment effect that consider the difference between
the average effect of treated and untreated individuals
are likely biased for the effect of interest \citep[][]{rubin1991aa, imbens2014causal}.

To make the study of causal effects concrete 
an analyst is required to 
define the treatment of interest, the outcome of interest that
she believes the treatment might affect, times at which
to conceptualize treatment happened, and times at which to measure the outcomes
\citep{splawa1990application,rubin1974estimating}. 
Within the context of a randomized experiment the
meaning and definition of these three components
are very straightforward, but in the context of an
observational study much greater care is required in 
defining them to allow for proper causal inference. In particular,
one might only observe a proxy for treatment and so the exact treatment
and, more importantly, the exact time of treatment
are unknown. When this is the case it becomes unclear when the
outcome of interest should have been measured in order to make 
a proper causal statement. This is the setting we consider in this paper. 

As a concrete example, we consider studying the relationship 
between the adverse
weather in February and March of 2015 across the 
United States and online battery searches at a major US retailer. The relationship between weather and sales has
been studied in the literature \citep[e.g., see][]
{murray2010effect,starr2000effects,zwebner2013temperature}, with
the sometimes implicit assumption of a clearly defined treatment (e.g., temperature, exposure to sunlight) and so causal statements are reserved for post-weather measurements.
In 
our case we are interested in studying whether the perception
of future extreme snow increases conversions but we do not observe this perception. Having observed the snow the methods proposed in Section \ref{sec:method} determines the time points prior to the snow event for which causal statements can be made.
In a novel data set,
provided by a large online advertising
technology firm,
we have measurements of these outcomes across February and March of 2015 for 79 designated marketing areas (DMA) 
across the United
States. For each of these super-metropolitan areas we have covariate information pertaining
to the demographics of the people living in the area, as well
as measurement of snow accumulation for each day. 




The rest of the paper is organized as follows: 
In Section~\ref{sec:notation} we briefly outline the formal notation
of the Rubin Causal Model and we introduce two new assumptions, which 
 complement the assumptions employed
in causal inference, that enable causal analyses of 
data that are missing an exact time of treatment. 
Section~\ref{sec:method} describes the proposed three stage 
methodology. Section~\ref{sec:sims} provides simulation results.
Details of the marketing 
data set and the data analysis are reported in Section~\ref{sec:data}.
Remarks follow, in Section~\ref{sec:disc}.


\section{A notion of plausible time of treatment}\label{sec:notation}

Throughout this paper we restrict ourselves to two levels
of treatment and for convenience we refer to ``control''
for the lower level of treatment, and  to ``treatment'' for
the higher. As in the classical causal inference literature $Z_{i}$ denotes
treatment indicator for an individual $i$ and is equal to 1 if
the unit is assigned to treatment at time $\Ti$
and is 0 otherwise \citep{imbens2014causal}.
Potential outcomes for each unit are denoted
by $Y_{i,\Ti,t}(Z_i=0)$ or $Y_{i,\Ti,t}(Z_i=1)$ for control and treated levels
of treatment when treatment occurs at time $\Ti$
and the outcomes are observed at time $t > \Ti$.

In contrast with the classical setting, the treatment indicator $Z_i$ and the 
time of treatment $\Ti$
remain hidden and instead, $D_i$ and $\toi$ are observed. 
$D_i$ is the indicator that an event associated with the treatment
occured at time $\toi$. In particular it is not necessary that
$Z_i=D_i$ and it is likely that $\Ti<\toi$. That is, the event
that is observed is not necessarily a perfect proxy for the 
actual treatment and the actual treatment time occurs
before the event. 
In what follows
we formulate a set of assumptions on the 
relationship between $Z_i$, $D_i$, $\Ti$ and $\toi$ that allow us to determine
the times $t$ for which an observed quantity $Y_{i,t}^\obs$ corresponds to the potential outcome under treatment or under control:
\begin{equation}\label{bigeq}
Y_{i,t}^\obs=D_i \cdot Y_{i,\Ti,t}(Z_i=1) + (1-D_i) \cdot Y_{i,\Ti,t}(Z_i=0)
\end{equation}
We then employ these assumptions to reconstruct an 
observational data set that is suited for causal inference. 

To develop these assumptions we require a new
notion that we refer to as the ``Last Plausible Randomized Experiment Time'' (termed
LaPRET and represented by $\Li$).
In the context of an idealized randomized experiment, outlined
in the left hand panel of Figure \ref{fig:ass2}, LaPRET defines the first time point at which
the two potential outcomes are differentiable. In other words,
in a randomized experiment where the treatment has an effect
after being administered at treatment time $\Ti$,
if the outcome were to be observed at time $\Oi\leq \Li$ then
no difference between $Y_{i,\Ti,\Oi}(Z_i=0)$ and $Y_{i,\Ti,\Oi}(Z_i=1)$ would be discernible (that is, a treatment effect would be undetectable).
On the other hand
observing at time $\Oi>\Li$ would yield a non-zero treatment
effect. It is clear that when there is no treatment effect, $\Li$ is not well defined -- the method proposed in this paper for identifying $\Li$ in observational studies accounts for this issue.

\subsection{Formal assumptions}

We require two new assumptions to discuss Equation \eqref{bigeq}. The first
facilitates the comparison between the realized treatment and the associated
event while the second provides a rationale for identifying the
times at which treatment could have happened.

\paragraph*{Assumption 1. (Unit Treatment Status Identification Strength)}

For treatment indicator $Z_i$ and associated event indicator
$D_i$ assume that

\begin{equation*}
{\rm cor}(Z_i, D_i) \geq 1 - \eta\ \forall i\text{ for }\eta\text{ small.}
\end{equation*}

\paragraph*{Assumption 2. (Constant Unit Time of Treatment)}

For LaPRET, $\Li$, treatment time $\Ti$ and associated event time $\toi$ assume
\begin{itemize}
	\item[(i)] $d_i =  \toi  - \Li =d >0$ \mbox{ for all $i$}
	\item[(ii)] $\Ti \leq \Li$  \mbox{ for all $i$}
\end{itemize}

The statement of Assumption 1 reflects potential uncertainty
about realized treatment status based on event status. In the extreme
setting where $\eta=0$ the assumption states that
$Z_i=D_i$ for all units $i$, fully revealing the treatment
indicator. This is the setting when a longitudinal 
study reveals information about a treatment that
occurred during a previous wave of the study. In the less
extreme setting of $\eta$ small, the assumption provides
a generative method for multiple imputed data sets all of which
are suitable for causal inference (see e.g. \cite{rubin1996multiple}). 
In particular, if we reconceptualize Equation~\ref{bigeq} as:
\begin{equation}
\Pr \bigm( Y_{i,t}^\obs=D_i, Y_{i,\Ti,t}(Z_i=1) + (1-D_i) \, Y_{i,\Ti,t}(Z_i=0) \bigm) ~\geq 1-\delta\label{eqn:reconc}
\end{equation}
 we have
$\delta=\delta(\eta)$ which
informs the interpretation of a sensitivity analysis that
varies $\eta$. When $\eta\neq 0$ there are multiple versions of the ``complete data'' as there is now a nonzero probability that $D_i=1$ and $Z_i=0$. This condition can easily be accommodated by constructing multiple data sets in which the correlation between $Z_i$ and $D_i$ is $1-\eta$ and reporting causal estimates across these data sets \citep{rubin1996multiple}.
\begin{figure}[b!]
	\centering
	\includegraphics[width=\textwidth]{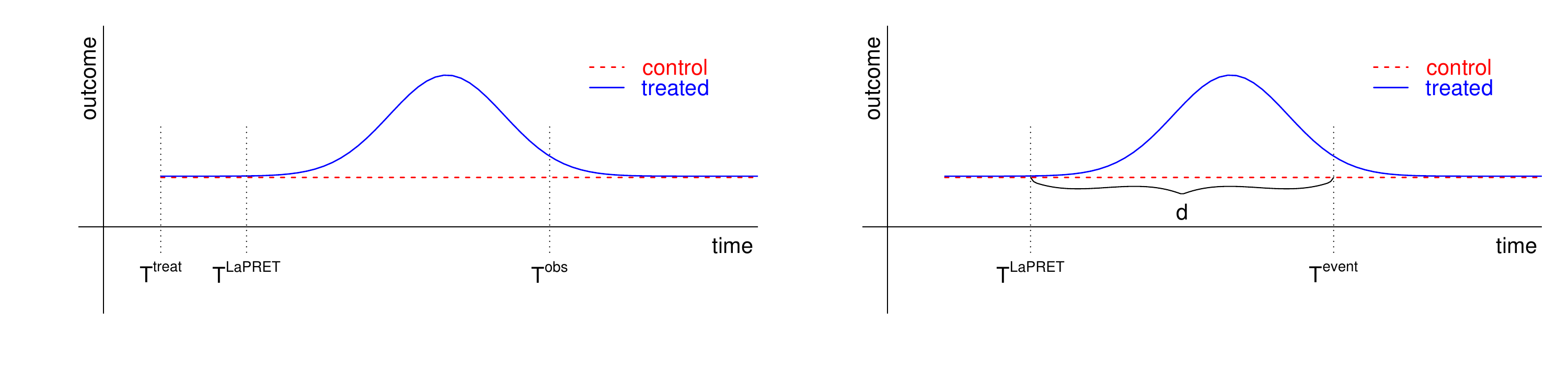}
	\caption{\onehalfspacing This is an illustration of Assumption 2. The left panel hand describes the scenario of an idealized experiment -- $\Ti$ is the known treatment time, $\Li$ is the LaPRET and $\Oi$ is the observation time. The right hand panel describes the observational data set where an event associated with treatment is observed at time $\toi$ and $d$ (of Assumption 2), the time between the event time and the LaPRET is the same for all $i$ (and hence does not require a subscript).}
	\label{fig:ass2}
\end{figure}
 
Assumption 2 is an identification assumption for LaPRET that explicitly addresses the lack of information about the time of treatment. The first part of the assumption states
that the time between an individual LaPRET and the time
of the associated event is equal for all individuals in the 
population. 
As such, by observing two similar individuals, one in the control
and one in the treatment groups (using the information
from Assumption 1), one can compute the LaPRET as illustrated in the right hand panel of Figure \ref{fig:ass2}.
The second part of the assumption states that the time
of treatment is always strictly prior to LaPRET. 
 meaning that 
observations at the LaPRET are potential outcomes and so
can be used to compute causal estimates.
Our methodology relies on the first part of this assumption - 
we estimate $\hat{d}$ from a pilot study and then leverage
this information in the larger population. If this assumption
is violated then the estimate of $\hat{d}$ is unreliable and 
care must be taken to insure estimates are causal. In such
situations, illustrated in the simulation study, a smaller
value of $d$ can be chosen to insure that the estimate
is causal. This suggests a relaxation of the condition in  part (i) of Assumption
2, to the get the following.

\paragraph*{Assumption 2'. (Stable Unit Time of Treatment)}

For LaPRET, $\Li$, treatment time $\Ti$ and associated event time $\toi$ assume
\begin{itemize}
	\item[(i)] $d_i =  \toi  - \Li >0$ \mbox{ for all $i$}
	\item[(ii)] $\Ti \leq \Li$  \mbox{ for all $i$}
\end{itemize}

\noindent In this more general case, there exists $d>0$ such that $d_i =  \Oi  - \Li>d$ for all $i$.

\section{Designing an observational study with unknown time of treatment}
\label{sec:method}

In this section, we describe how one may leverage
the assumptions of Section \ref{sec:notation} in order to take a large data set that does
not include treatment indicators or time of treatment indicators
and to produce a potentially reduced data set that can be
used for conceptualizing and performing a complete observational study \citep{Rosenbaum:2002aa,Rosenbaum:2010aa}.

\subsection{Conceptualization of treatment}

Unlike in classical causal inference where treatment and treatment
time are known to the scientist, in this setting a proper notion of treatment
must be conceptualized. The issue here is similar to the one faced in the 
perceived treatment literature where one cannot use race or sex 
as the treatment in an experiment but instead ``perceived race'' or ``perceived
sex'' can be used \citep{greiner2011causal}. We also rely on the notion of 
``perceived treatment''. For example, in the 
application to the advertising data in Section \ref{sec:data}
the treatment is the expectation of an extreme snow
event happening in the future perceived by individuals. The notion of perception
is needed here as the true treatment is never observed in our
setting and so we reconstruct it based on an event that would
have been foreshadowed by such a perception. This is exactly the
type of correlation between the treatment indicator $Z_i$
and the event indicator $D_i$ that Assumption 1 describes quantitatively.
It must be noted that $D_i$ identifies both the treatment
and the control groups and so it is likely that some units
will be discarded in order to better satisfy Assumption 1.
This data reduction step is explored in detail in Section \ref{sec:data} where
the only units allowed to be considered as control units
are those that always experience less than a thresholded amount of snow. 

Once we have identified the event that serves as proxy for the true 
treatment we must identify the correlation in Assumption 1. In a longitudinal study setting,
where a later wave question (serving as the indicator $D_i$) might
ask if an individual received an intervention at a 
previous wave (the treatment indicator $Z_i$) the correlation 
can be set to one. In more ambiguous situations, such
as the one discussed in the Section \ref{sec:sims}, a sensitivity
analysis based on this correlation should be performed.

\subsection{Pilot study to identify plausible times of treatment}

Once the treatment is defined and a variable $D_i$ is identified
we need to find the individual LaPRET values $\Li$. Since these
are only identifiable from the joint distribution of outcomes we
must infer those from the data. However, if we use a data driven
approach that considers the complete data set we would be violating
a fundamental principle of causal inference that does not allow parameters
in the analysis to depend on the post-treatment data. 

To overcome this difficulty we propose to perform a pilot study to
identify $\Li$. An example of this approach was recently undertaken by \citet{wager2015estimation} to apply decision tree methods to causal inference. Under the pilot study we consider a small sample
of  individuals that are identified as treated and as control units. 
Once this sample is chosen we match treated and control pairs. The choice of a particular matching mechanism depends on the applied problem and the available observed covariates associated with each unit -- in our simulations and applied example we employ propensity score matching \citep{rosenbaum1983assessing}. In a slight abuse of notation, the matched pairs are now identified by the subscript $i$.
Leveraging Assumption 2(i) (or 2(i)$^\prime$) we can identify 
the difference between the time of the event $D_i$ and
the LaPRET, $d_i=\toi-\Li$, as it is assumed to be constant (or greater
than a non-zero constant) for all units. 
As such finding
$d_i$ becomes a problem of identifying time points $t$
where $Y_{i,t}^\obs|D_i=1$ and $Y_{i,t}^\obs|D_i=0$ are close. Letting $\Delta_{i,t}$ equal that difference for matched pair $i$ at time $t$
we say that the LaPRET for matched pair $i$ is 
\begin{align}
\Lihat \in \argmax_{t<\toi}\Bigm\{ t 
 & \text{ s.t. } |\Delta_{i,t}|<\max_t \frac{|\Delta_{i,t}|}{\alpha}\text{ and } \nonumber \\
 & \text{ s.t. } |\partial\Delta_{i,t}|<\epsilon, ~|\partial\Delta_{i,s_1}|>\epsilon, ~|\partial\Delta_{i,s_2}|<\epsilon \text{ for some } s_2>s_1>t \Bigm\}. \nonumber 
\end{align}
That is, the estimated LaPRET for pair $i$ is at the 
maximal time point such that the difference $\Delta_{i,t}$
is smaller than a $(1/\alpha)$ fraction of the maximal difference
and the rate of change of $\Delta_{i,t}$ with respect to time is small but there 
exists a later time point where the rate of change is large.
The parameter $\alpha$ captures the expected variability in the
values of $Y_{i,t}^\obs$. Small values of $\alpha$ allow for larger differences between treated and control observations to be evaluated as no treatment effect. As such, larger values of $\alpha$ lead to more conservative $\Lihat$ -- those that are closer to $\toi$.
The parameter $\epsilon$
controls the rate of change of $\Delta_{i,t}$. It insures
that the procedure is able to differentiate between 
no effect (where $Y_{i,t}(1)=Y_{i,t}(0)\ \forall t$) and 
a situation where the effect of treatment either induces
volatility or decreases to zero before the time
of observation. Small values of $\epsilon$ require the volatility between the treated and control observations to be small almost all the time while larger values allow for lots of volatility but require that a bigger volatility event occurs. As such both too small and too large values of $\epsilon$ lead to extremely conservative behavior. 
We study the behavior of $\alpha$ and $\epsilon$ in a 
simulation study.

After computing $\Lihat$ we can construct $\hat{d}$ for
Assumption 2 using some function of the set $\{\toi-\Lihat \}_i$.
In the simulation study we explore using the average of those differences 
to choose $\hat{d}$. Choosing the minimum forms
a conservative estimate of $d$ that 
accommodates the relaxed Assumption 2$^\prime$.

In practice, the pilot study can be performed on a subset of 
units treated at a period of time before or after the main study happened.

\subsection{Main study}

The final step of the pipeline is the construction
of a possibly reduced data set from the units that were not
used in the pilot study which allows for valid causal inference \citep{Rosenbaum:2002aa,Rosenbaum:2010aa,imbens2014causal}.
In particular, having identified the relationship
between the treatment indicator $Z_i$ and the
event indicator $D_i$ as well as the latency between
LaPRET and the event in the previous two steps one constructs
the data set as follows: (1) among the remaining units, construct
a matched sample then (2) for each matched pair, discard information 
that is recorded prior to $\Lihat=\toi-\hat{d}$. The remaining
data then represents the potential outcomes under treatment
and control.



\section{Simulations}\label{sec:sims}


The purpose of our method is to give the analyst an interval of
days $\hat{d}$ prior to the event, for which he can make causal
statements about the conceptualized treatment. This section has two objectives: first, we assess how
the choice of parameters $\alpha$ and $\epsilon$ affects the behavior
of our method in different scenarios. Second, we explore how this 
behavior is affected by different types and levels of noise in the observations.

\subsection{Design choices} \label{sect:sims-description}

\begin{figure}[htbp]
	\centering
	\includegraphics[width=0.95\textwidth]{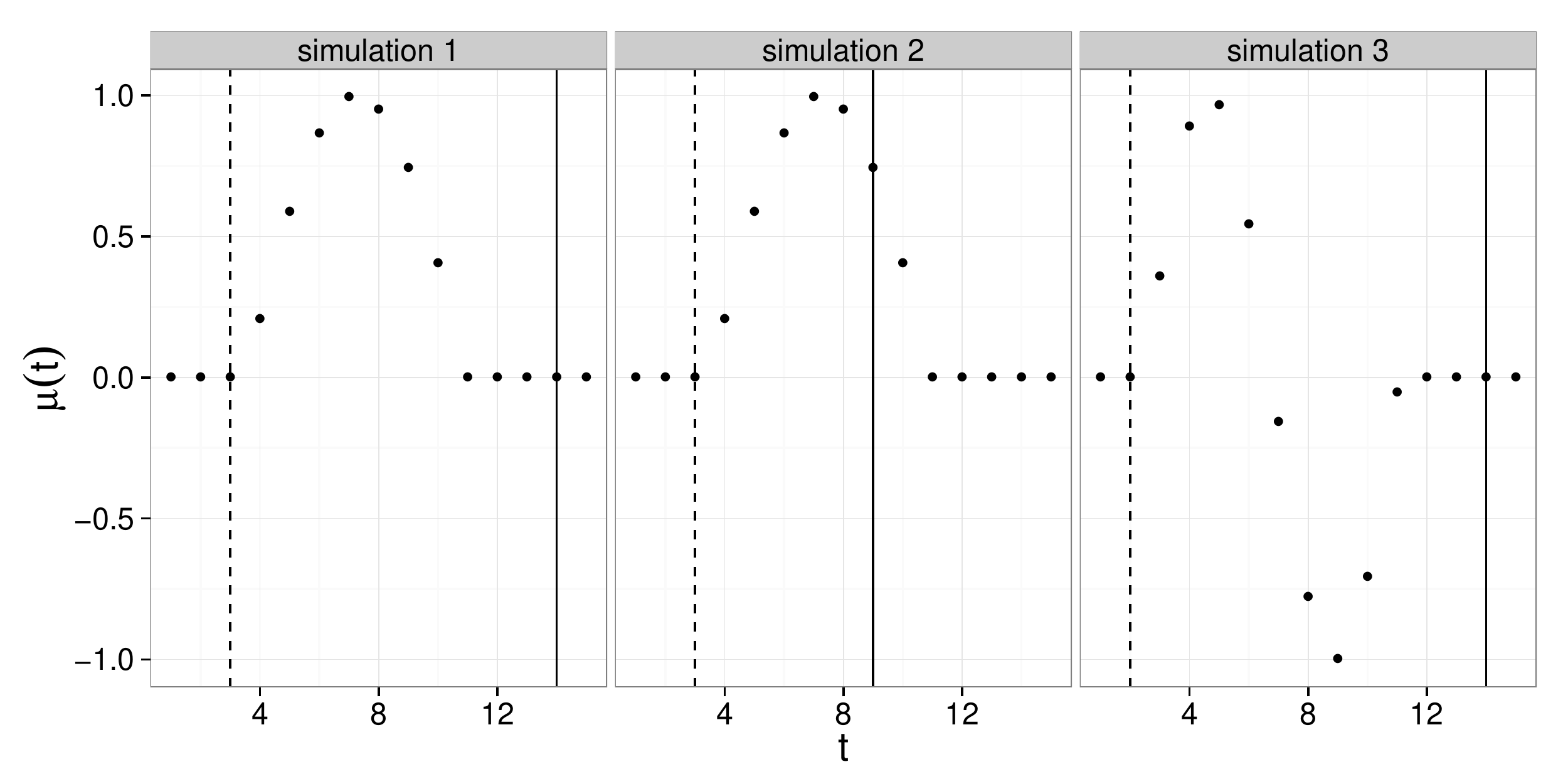}
	\caption{\onehalfspacing Response surface for the three simulations. The solid lines are the 
	observation times, $\toi$, and the dotted lines are the LaPRET, $\Li$}
	\label{fig:resp}
\end{figure}

We conduct three experiments of increasing complexity, trying capturing different 
real life response profiles. For each of the three scenarios, we consider a pair of ``idealized
response surfaces'' under control and treatment -- that is, 
response surfaces with no noise that capture one of the three 
scenarios of interest. We denote pair of responses corresponding
to the $k^{th}$ scenario by $(\mu^{(k)}_0(t), \mu^{(k)}_1(t))$. For
all three scenarios, the idealized control response surface is the 
flat line at zero, $\mu^{(k)}_0(t) = 0$. The idealized response 
surfaces are represented in Figure~\ref{fig:resp}.

We also consider two potential sources of noise in the observations. 
First we simulate the fact the potential outcomes are only noisy versions of the idealized
response surface. So for each scenario $k$, we generate $N$ treatment and 
control potential outcomes surfaces:
\begin{equation}
	Y^{(k)}_{i,t}(a) \sim \mbox{Normal }(\mu^{(k)}_a(t), \sigma^2)
\end{equation}
for $a=0,1$, and $i=1\ldots N$. The noise parameter $\sigma$ was given the following
values $\sigma = 0.005, 0.01, 0.015, 0.02$, but such that the potential outcome surfaces generated
in a given simulation all shared the same parameter $\sigma$. The second source of 
noise we consider is with the time of observation $\toi$ of the event, which we simulate
with a contamination model: $\toi = \Li + d + c$ where $\Li$ and $d$ are fixed and $c$
is a contamination model that is governed by one of the following distributions:
 \begin{align*}
 & f_1(c) = 0 \text{ w.p. 1} 
 & \qquad
 f_2(c) = \begin{cases}
 	-1 & \text{ w.p. 0.25}\\
	0  & \text{ w.p. 0.5}\\
	1  & \text{ w.p. 0.25}
	\end{cases}%
 \\
 & f_3(c) = \begin{cases}
 	-1 & \text{ w.p. 0.1}\\
	0  & \text{ w.p. 0.5}\\
	1  & \text{ w.p. 0.4}
	\end{cases}%
 & \qquad
 f_4(c) = \begin{cases}
 	-1 & \text{ w.p. 0.4}\\
	0  & \text{ w.p. 0.5}\\
	1  & \text{ w.p. 0.1}
	\end{cases}	
\end{align*}

For each of the $4\times 4=16$ possible noise structures $(\sigma^2,f_i)$, 
we study the behavior of our method for $\alpha = 1, 6, \ldots 96$ and 
$\epsilon = 0.0001, 0.02, 0.2, 0.3, 0.4, 0.5$. So for each combination of parameters
$(\sigma^2, f_i, \alpha, \epsilon)$, we simulated a data set with $N=600$
observations, and compute $\hat{d}$ as in \textit{step 2} of Section~\ref{sec:method}. We
forgo matching in the simulations since this is beyond the scope of our contributions. The 
three simulations are described in more details below, and
the results are analyzed in Section~\ref{section:analysis-simul-results}.

\subsubsection{Non-negative effect curve, zero at observation time}
\label{sec:sim1}

The first simulation, we consider the scenario in the left panel of 
Figure~\ref{fig:resp}, where the LaPRET is $\Li = 3$, the event is 
observed at time $\toi=14$ (and so $d_i = \toi-\Li = 11$) and the treatment response surface 
is represented, for mathematical convenience, by the function 
$\mu^{(1)}_1(t)= \max \bigm\{ 0, ~\sin(\frac{2\pi}{15}(t-3.5)) \bigm\}$. In this 
setup, the true effect is always non-negative, but is zero at 
the observation time, and so a naive analysis might find no 
effect even when one is present. This version of a response
is likely when the natural event is catastrophic---if individuals
expect a snow storm, for instance, they might order batteries and snow tires in advance so 
as to have them ready for after the storm.


\subsubsection{Non-negative effect curve, positive at observation time} 
\label{sec:sim2}

The second simulation considers the same response surface
as Simulation~1 but a shifted observation time $\toi = 9$ such
that $d=6$ as displayed in the second panel of Figure~\ref{fig:resp}.
In this scenario, the true effect is still always non-negative, but 
the observation time corresponds to a point where there is still
a difference between the treated and control levels of the 
potential outcomes. This is a possible response surface
for insurance quote requests due to an upcoming
storm---there is an increase due to the forecast
and it does not necessarily go back down to the
previous levels until after the storm. 
Similar response surfaces were observed
in for online marketing 
campaigns \citep{lewis2011here}.

\subsubsection{Positive-negative effect curve, zero at observation time}
\label{sec:sim3}

The third simulation introduces volatility into the 
response surface. Here $\mu^{(k)}_1(t)$ is as in \eqref{sim3}, $\Li=2$ and
$\toi=14$. In this scenario, the true effect is positive immediately 
after $\Li$, but then changes sign before reaching zero shortly before $\toi$. 
In particular, there is a point of zero effect between $\Li$ and $\toi$ which does
not correspond to the LaPRET. In the context of medical trials, this volatility 
could correspond to the side-effect of a drug on a person's blood pressure. It could
also correspond to the purchasing of commodities before a big storm -- the dip 
illustrated in the right panel of Figure~\ref{fig:resp} corresponds to the fact that once
an individual stocks up on a commodity, he is likely to buy less of it for a while. Another
interpretation is that in anticipation of the event, individuals move their usual purchasing
day to before the event, provoking a dip in the days immediately preceding the event.
\begin{equation}\label{sim3}
	u(t) = 
	\begin{cases}
		0 & \text{if $v(t) \leq 0$ and $t < 4$}\text{ or }
		\text{if $v(t) \geq 0$ and $t > 10$}\\
		\sin(\frac{3.5\pi}{15}(t-2.5)) &\text{otherwise}.
	\end{cases}
\end{equation}

\subsection{Analysis of the results}
\label{section:analysis-simul-results}

\begin{figure}[t!]
	\centering
	\includegraphics[width=0.95\textwidth]{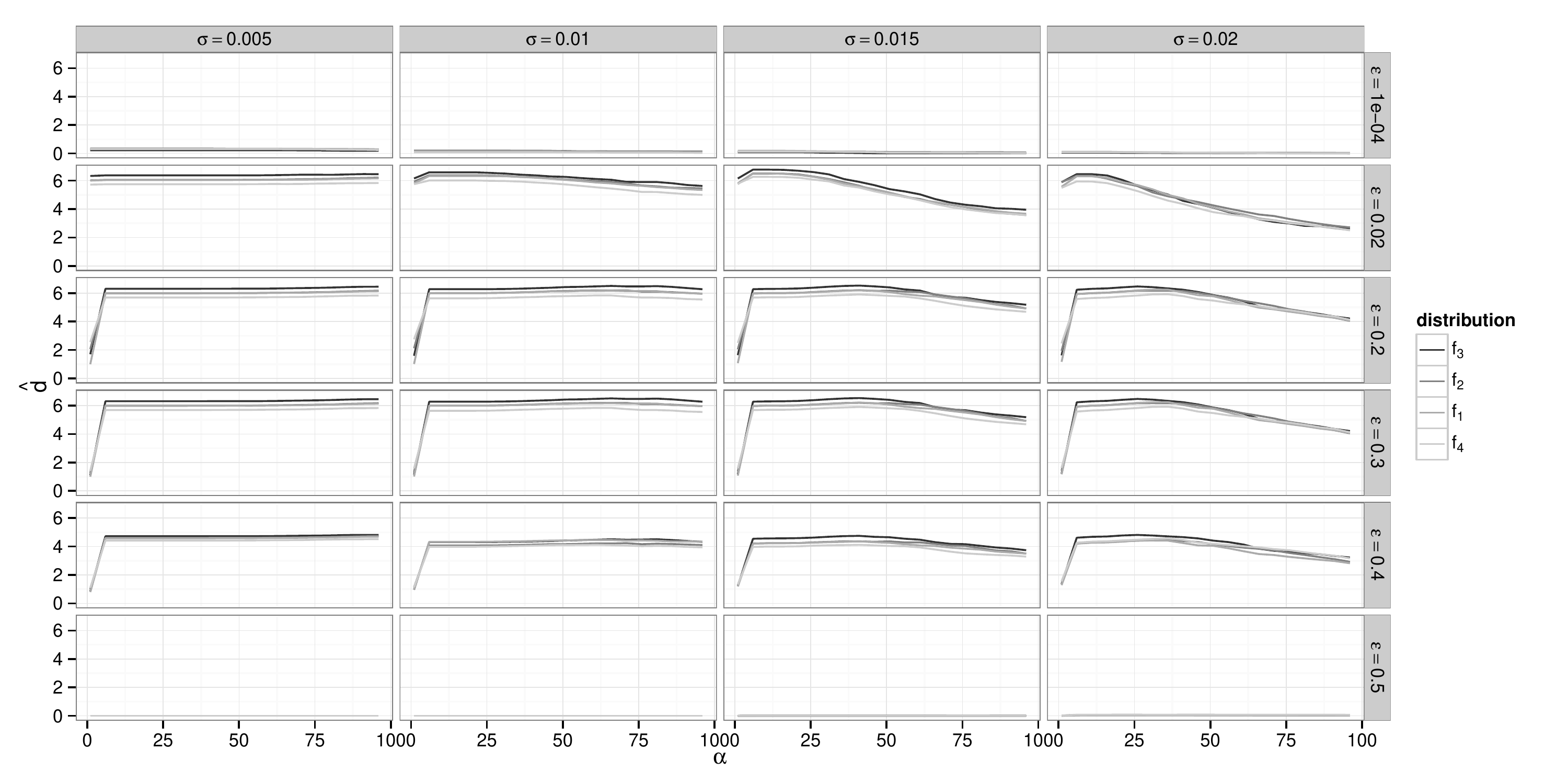}
	\caption{\onehalfspacing Estimate $\hat{d}$ as a function $\alpha$ under different
	contamination models, different levels of noise, and different values of $\epsilon$ for simulation 2, described in Section \ref{sec:sim2}.}
	\label{fig:simtest2}
\end{figure}

\begin{figure}[t!]
	\centering
	\includegraphics[width=0.95\textwidth]{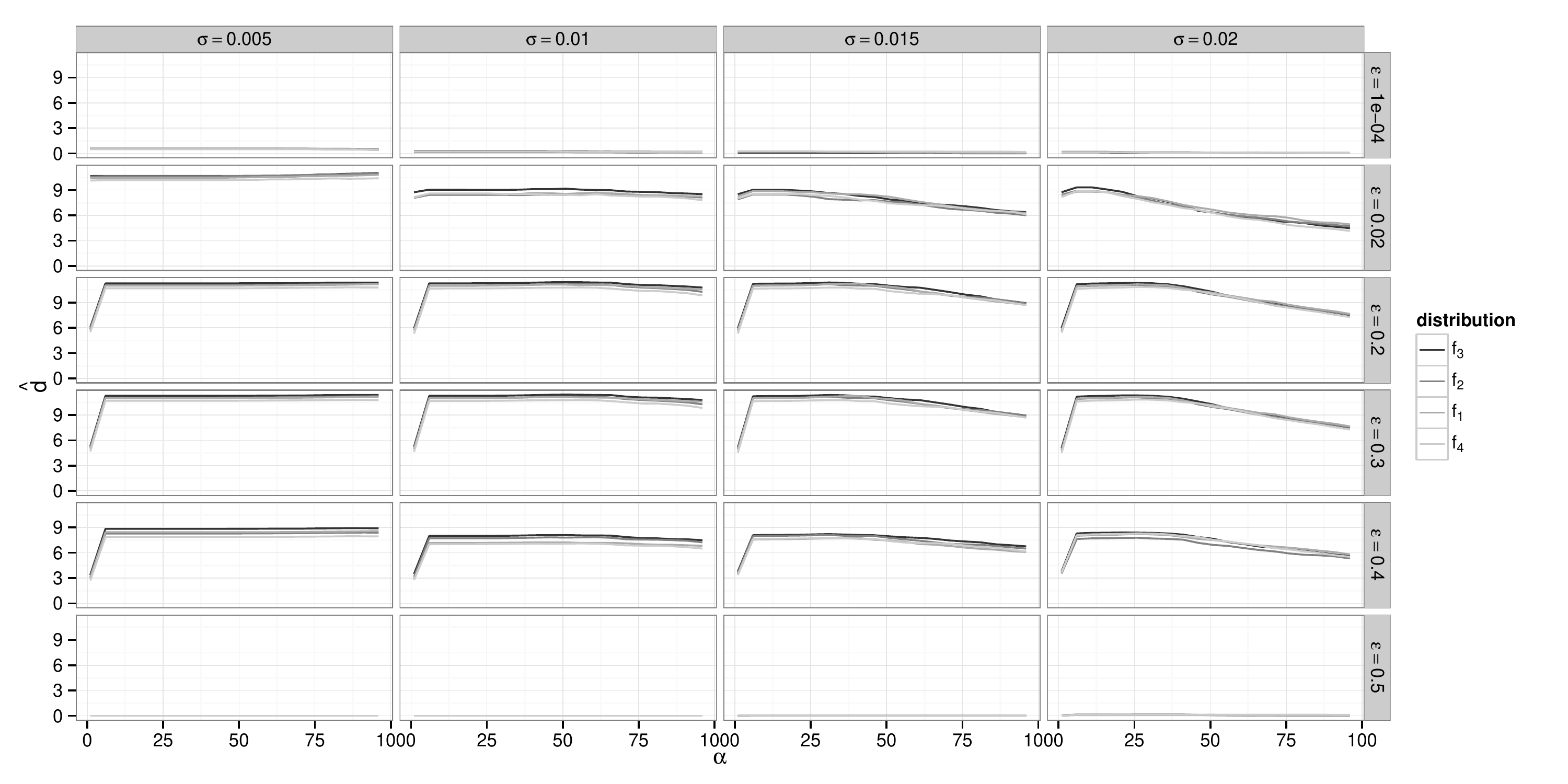}
	\caption{\onehalfspacing Estimate $\hat{d}$ as a function $\alpha$ under different
	contamination models, different levels of noise, and different values of $\epsilon$ for simulation 1, described in Section \ref{sec:sim1}.}
	\label{fig:simest1}
\end{figure}

\begin{figure}[t!]
	\centering
	\includegraphics[width=0.95\textwidth]{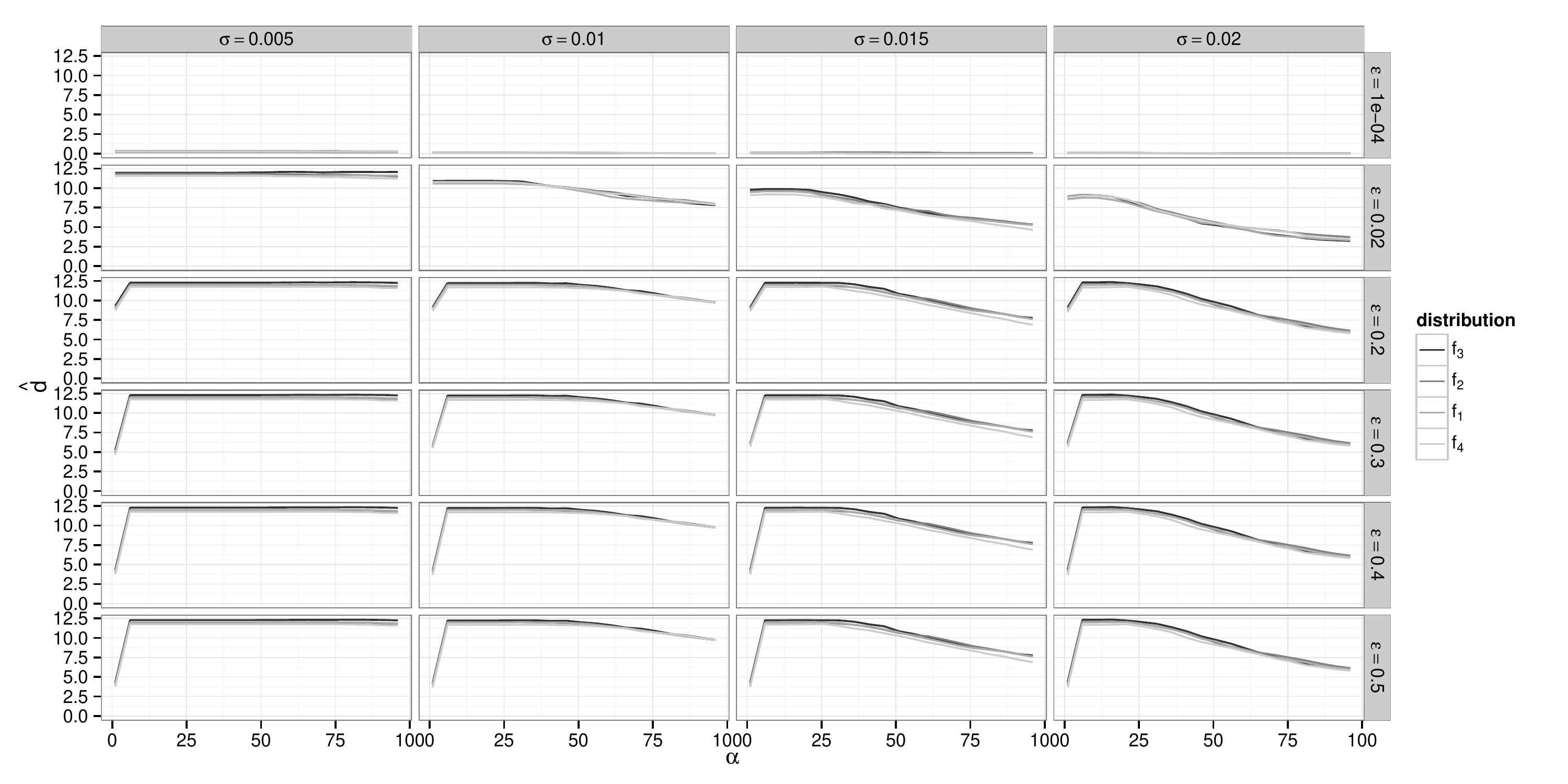}
	\caption{\onehalfspacing Estimate $\hat{d}$ as a function $\alpha$ under different
	contamination models, different levels of noise, and different values of $\epsilon$ for simulation 3, described in Section \ref{sec:sim3}.}
	\label{fig:simest3}
\end{figure}

The effect of the different parameters is similar in all three scenarios, so we
only describe the results of the second Simulation, described in Section \ref{sec:sim2}. The results are summarized in Figure~\ref{fig:simtest2}. 
Figures \ref{fig:simest1} and \ref{fig:simest3} summarize the results of the first and third simulations, respectively. We start by noticing
that although the contamination models have a small impact on the aggregated value of $\hat{d}$,
this can have a large impact on the integer part of $\hat{d}$. This quantity, which we denote by $\floor{\hat{d}}$, is the quantity of interest
since it represents the number of days before the event date for which we can make causal
statements. For instance in Figure~\ref{fig:simtest2}, for low levels of $\alpha$, and $\sigma=0.005$,
we see that $\floor{\hat{d}} = 6$ for contamination models $f_1, f_2$ and $f_3$, but $\floor{\hat{d}} = 5$ 
for $f_4$. This uncertainty is difficult to account for, and our method is very sensitive to it. For
fixed values of $\sigma$ and $\epsilon$, we see that $\hat{d}$ is close to the true value $d$ for
low values of $\alpha$, but then decreases as $\alpha$ increases. For very low values of $\alpha$,
however, the $\hat{d}$ decreases. For fixed values of $\alpha$ and
$\sigma$, $\hat{d}$ is close to zero for low values of $\epsilon$, increases until a certain point with $\epsilon$,
then decreases again to reach zero for high values of $\epsilon$. Finally, we see that as the noise
level $\sigma$ increases, $\hat{d}$ decreases, especially for high values of $\alpha$.\\


A decrease in the value of $\hat{d}$ is a conservative behavior from a causal perspective, as 
it reduces the range of effects that we can call causal. This means, based on our observations in 
the previous paragraph, that as the noise $\sigma$ increases, the estimator becomes increasingly
conservative in these scenarios, which is a desirable behavior. This property of our estimator is 
further explored in Section~\ref{sec:data}. Our observations have also made clear the fact that 
increasing the value of the parameters $\alpha$ tends to make the estimator more conservative.

\section{Analyzing the effect of snowfall on online behavior}
\label{sec:data}

\begin{figure}[!htbp]
	\centering
	\includegraphics[scale = 0.3]{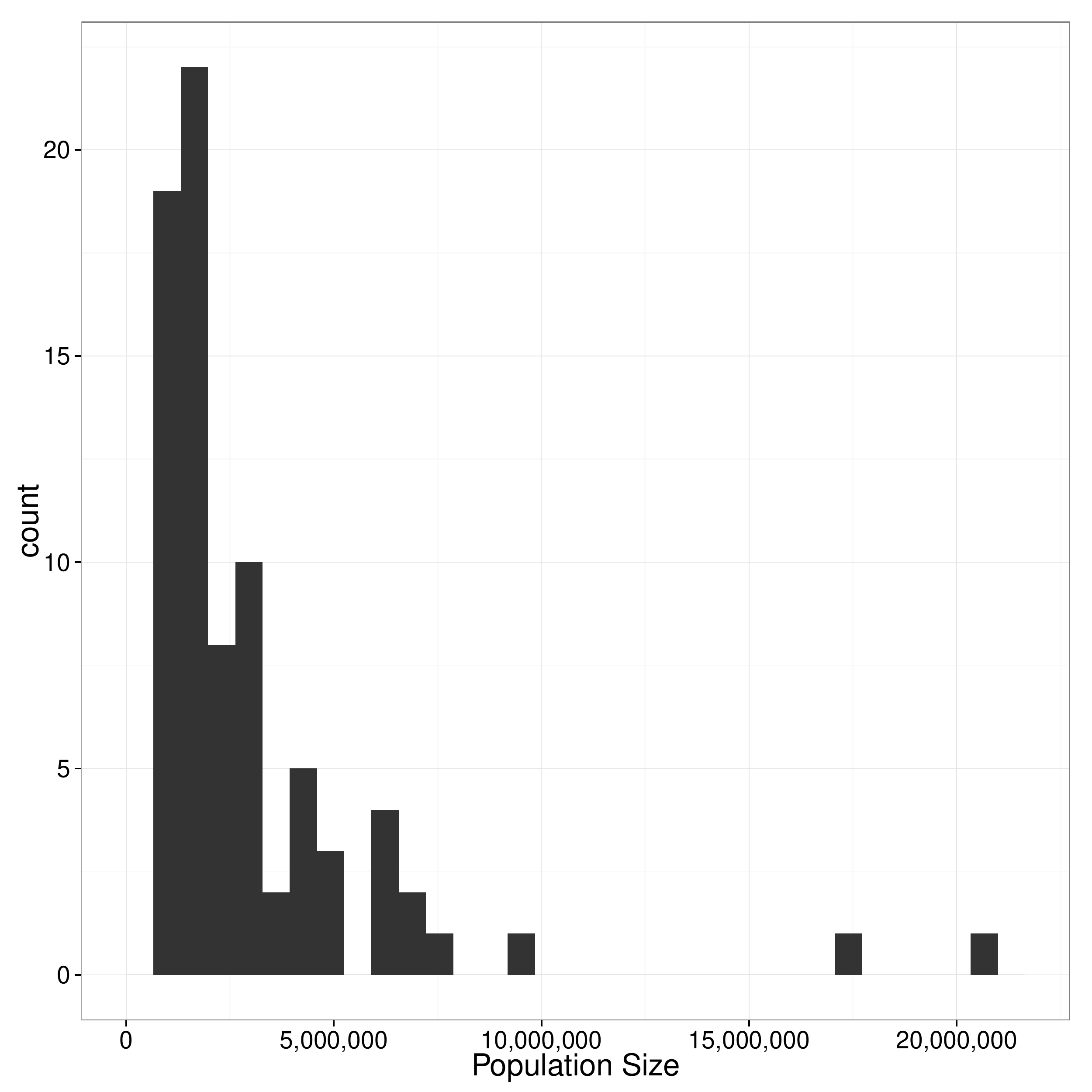}
	\caption{Distribution of the population of the 79 DMAs considered.}
	\label{fig:datapop}
\end{figure}

\begin{figure}[htbp]
	\centering
	\includegraphics[scale = 0.3]{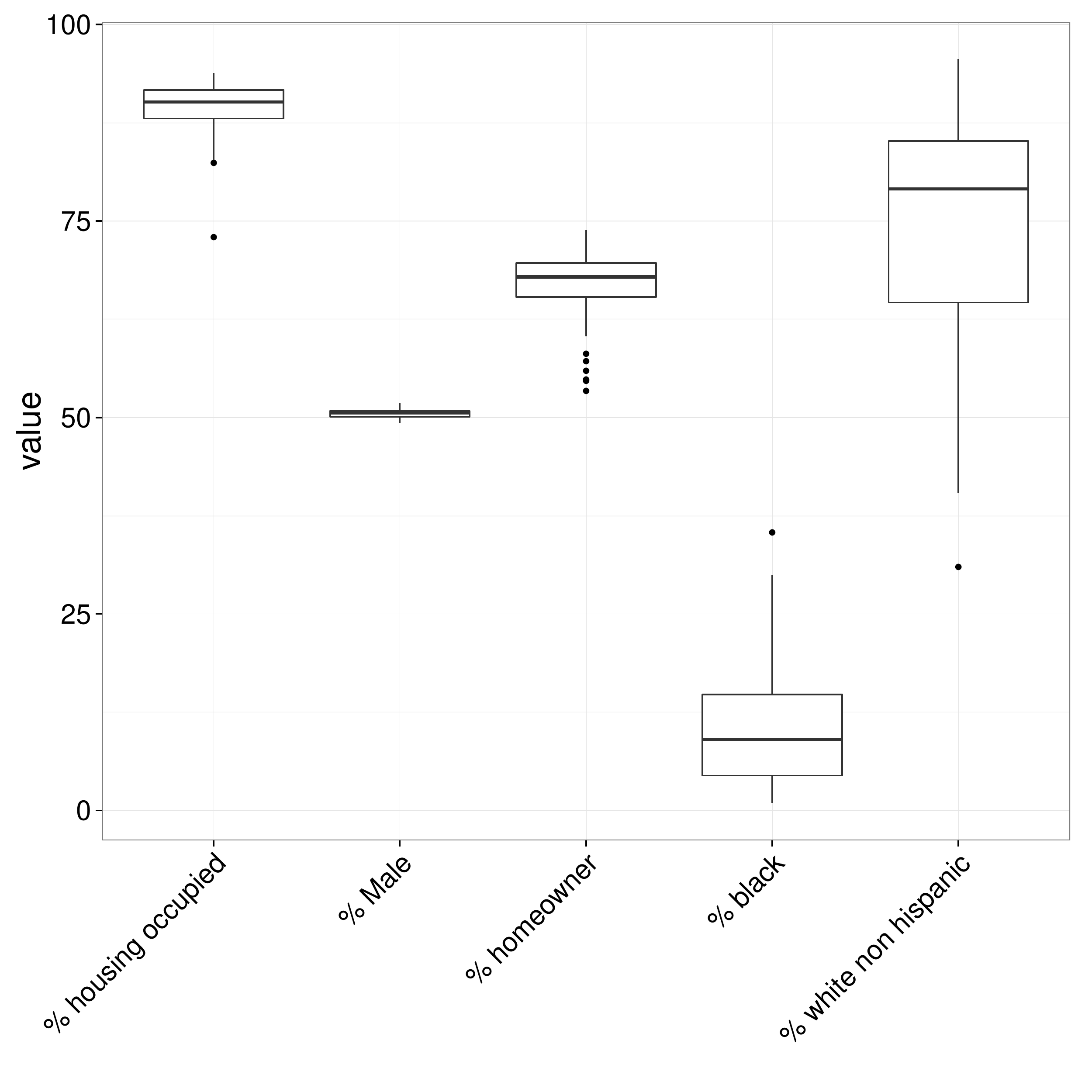}
	\caption{Distribution of the demographic variables across
		the 79 DMAs considered}
	\label{fig:datasum}
\end{figure}

In this section we estimate the causal treatment effect of
perceived large quantities of future snow on sales of products
on the internet. Our estimates are based on data  
provided by MaxPoint Interactive Inc., an advertising
technology company based in Raleigh, North Carolina. The data was 
provided according to designated marketing area (DMA) which 
corresponds to 79 super-metropolitan areas in the United States. For 
each DMA, the data contains three types of information over a period of
two months: 
 (1) searches for batteries on a major retailer's website, which we will consider as the outcome of our analysis, 
 (2) Demographic information from the census bureau, summarized in Figure~\ref{fig:datapop} and Figure~\ref{fig:datasum}, and 
 (3) the cumulative daily snowfalls. This data set is interesting as it can be seen as recording a natural experiment at the nationwide scale, in which the treatment time is unknown \citep[e.g., see][]{angrist2000aa,dunning_natural_2012,phan2015aa}.
%

\subsection{The data}
\label{section:synthetic-data}

Due to the large populations and geographical extents that
different DMAs cover, it is unreasonable to believe that an 
aggregate of weather at the DMA level will constitute a treatment
for all units in said DMA. Another way to say this is that DMAs
are not homogenous when it comes to weather, and there
can be a significant weather event happening in a DMA that will
affect only a small portion of the units inside that DMA. We thus
build a synthetic scenario that better illustrates the methodology
we propose. We introduce the concept of tradezones which can be
thought of as smaller subergions of DMAs, within which weather will
be more homogeneous. For each DMA, a number of tradezones proportional
to its population, totaling 3676 synthetic tradezones over all DMAs. 

For tradezone $j$ in DMA $i$, we create for each day $t$ a synthetic 
observation from a $\text{Normal }(\mu_i(t), \sigma^2)$ for $y_{ij}(t)$ truncated below at zero, 
where $\mu_i(t)$ is the real observed outcome for DMA $i$ on day $t$, such that
$y_{ij}(t) \geq 0$ for all $t$. Varying the noise level $\sigma$ allows us to
evaluate the robustness of our procedure. We ran simulations for $\sigma = 2^k$, $k= 1 \ldots 7$.
To give some perspective about the relative size of the noise, we report that the median 
outcome in the data across all DMAs and all days is 13, and the $95^{th}$ percentile is $60$.

Another element required to construct a realistic data set is to handle the 
weather, keeping in mind that the original motivation for introducing the
tradezones was to deal with weather heterogeneity within DMAs. To address
this, we set a threshold $h=1 \,\kg/\m^2$ of snow, and for each DMA, we considered
the days for which the snow precipitations exceeded that threshold. Suppose
there where $K$ such days for DMA $i$, which we will label $\{t_1, \ldots, t_K\}$, and
let $S_1, \ldots, S_K$ be the corresponding precipitations in $\kg/\m^2$. For each 
tradezone $j$, we selected a day of observation $T^{({\rm event}, j)}$ at random among
$\{t_1, \ldots, t_K\}$, such that $p(T^{({\rm event},j)} = t_l) = \frac{S_l}{\sum_k S_k}$, that is,
proportional to snow precipitations in the eligible days. Tradezones in DMAs for 
which no snow precipitation exceeded $l = 0.3 \,\kg/\m^2$ were all assigned to the 
control group, while all remaining tradezones were ignored.

With this definition of treatment, it is reasonable to say that in most cases,
no unit with $D_i = 0$ would have had the perception that it is going to be hit
by an event such as $1\, \kg/\m^2$ of snow. This definition justifies setting the correlation in Assumption 1 to be one (that is $\eta=0$) and hence the observed outcomes are realizations of the potential outcomes (that is $\delta=0$ in Eq~\eqref{eqn:reconc}). 
Larger $h$ strengthen this assumption but drastically reduce the number of DMAs in treatment.  


Census data at the DMA level are used to complete the synthetic data set. For each DMA $i$, we have a vector of covariates $x_i$.
Since the propensity score matching is performed at the tradezone level, we let
tradezone $j$ in DMA $i$ inherit its covariate vector $x_i^{(j)}$ from the 
parent DMA. That is, we assume that $x_i^{(j)} = x_i\ \forall j$.

In summary, for each DMA (for which we have real data), we have generated
synthetic tradezones, for which we simulated synthetic observations, and
selected a day of observation for the snow event based on the distribution of snowfall in the DMA. Our purpose is
to illustrate how an analyst would apply our method to that kind of data, 
and what kind of robustness he should be expecting. The results we report below represent causal estimates of nationwide battery searches online 
(for a major american retailer) based on the synthetic data.

\subsection{Sensitivity analysis}

For each of the seven levels of noise $\sigma = 2^k$, $k=1\ldots7$, we generated 
one data set as described in Section~\ref{section:synthetic-data}, and ran both a 
pilot study and an analysis as described in Section~\ref{sec:method}, with parameters 
$\alpha = 2.5$ and $\epsilon = 4.$. Injecting different levels of noise assesses the 
sensitivity of the analysis to the synthetic data generating process. We discuss heuristics
for the choice of $\alpha$ and $\epsilon$ in Section~\ref{section:heuristics}.

\begin{figure}[b!]
	\centering
	\includegraphics[width=0.95\textwidth]{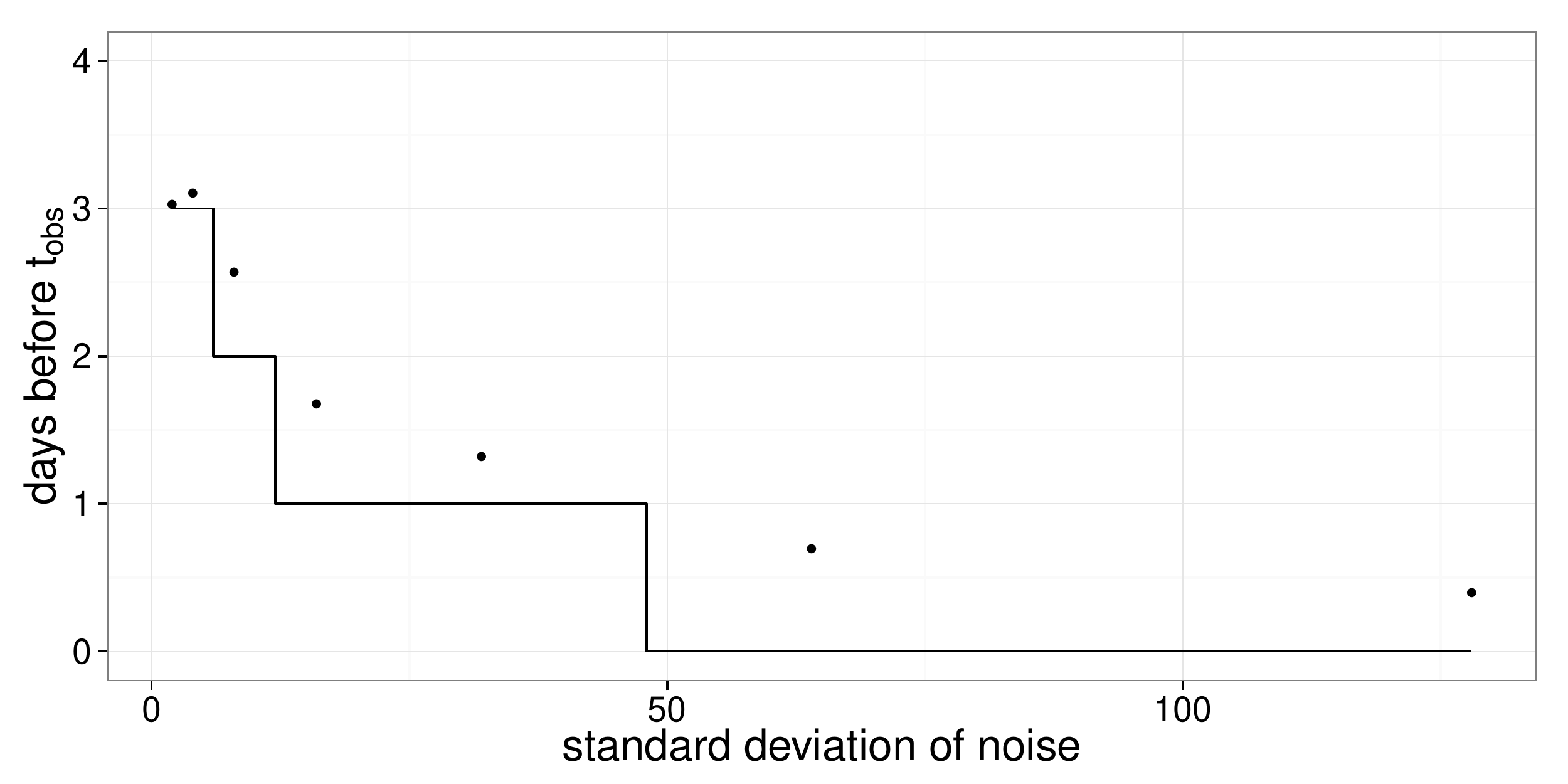}
	\caption{Sensitivity of $\hat{d}$ to noise in the tradezones}
	\label{fig:dhat_sensitivity}
\end{figure}

To stabilize the data and account for
difference in baseline outcomes among DMAs, we consider lagged differences between outcomes. That is, if $y_{ij}(t)$
is the outcome of tradezone $j$ in DMA $i$ at time $t$, then we considered the transformed
outcomes $y^*_{ij}(t) = y_{ij}(t) - y_{ij}(t-1)$ (we ignore the first day in the data set), and 
applied the method in Section~\ref{section:synthetic-data} to the transformed data.This analysis provides insight into the causal changes in behavior from day to day due to perceived future snow.
Figure~\ref{fig:dhat_sensitivity} shows the values of $\hat{d}$ obtained in the seven pilot studies,
where each pilot study contains 878 of the 3676 tradezones. The solid line shows the integer part of $\hat{d}$ which is the real
quantity of interest, since $d$ is the number of days before the event for which we can report
the effect as being causal. We see that the $\hat{d}$ (and hence the solid line) decreases
as the standard deviation of the noise increases in our simulations. This confirms the 
conservative behavior identified in Section~\ref{sec:sims}: the noisier the data, the more
conservative we become about making causal statements.

\begin{figure}[t!]
	\centering
	\includegraphics[width=0.95\textwidth]{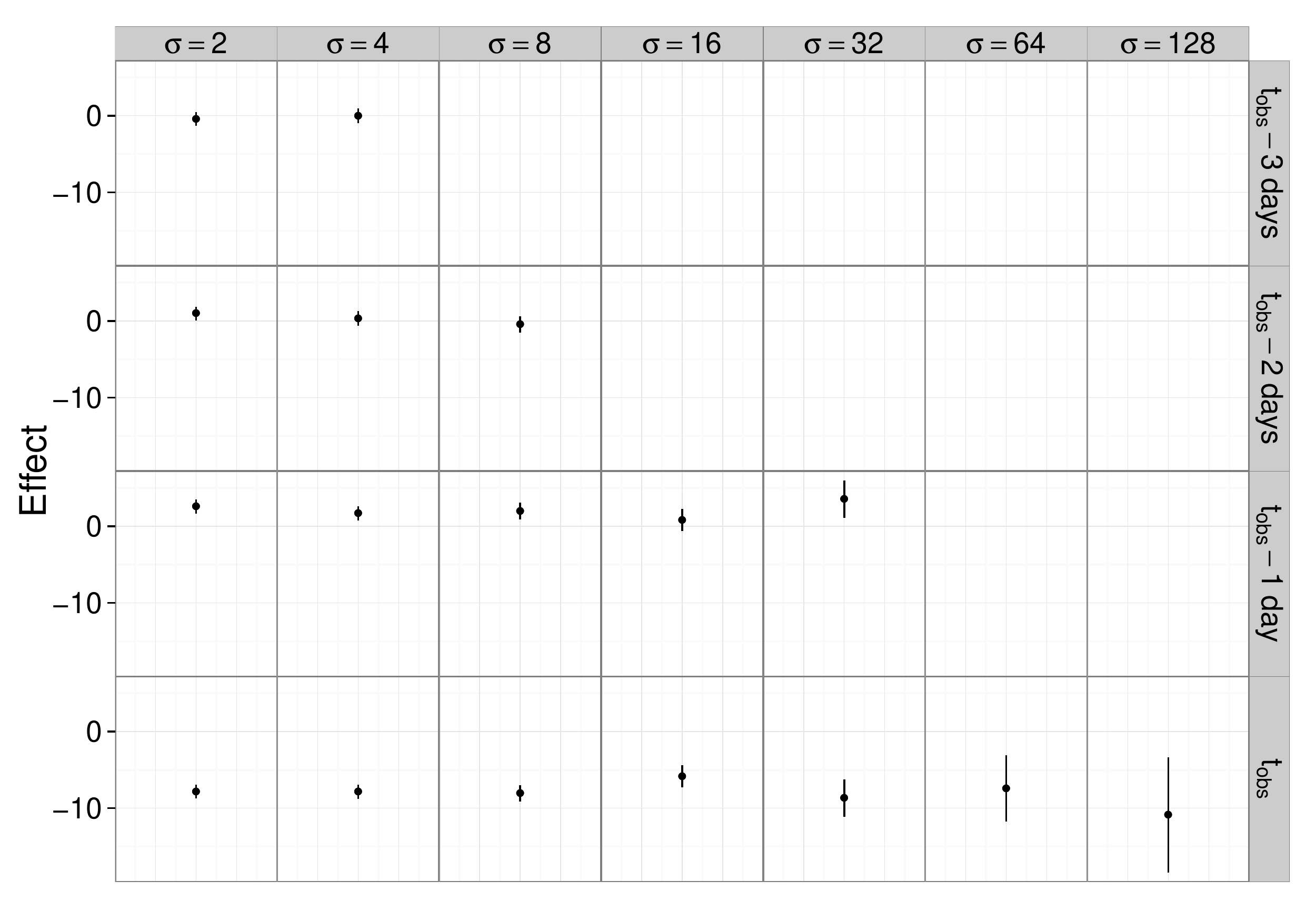}
	\caption{\onehalfspacing Sensitivity of the estimated effect to noise in the tradezones. Each frame represents the average causal effect and a 95\% confidence interval.}
	\label{fig:effects_sensitivity}
\end{figure}

After obtaining values for $\hat{d}$ from the pilot studies, we completed the analysis
for each data set based on the remaining 2798 tradezones. Figure~\ref{fig:effects_sensitivity}
summarizes the results, and illustrate the conservative nature of our procedure: for 
high levels of noise, the only effect that can be reported as causal is that on the day
of the observed event. For low values of the noise however, causal statements can 
be made up to two days prior to the day of the weather event.

From this data we conclude that there is a causal relationship between searches for batteries and perceived future snow events that takes on the form of the ATE in Simulation 3 (see Fig~\ref{fig:resp}). That is, the causal effect appears to be consistently positive and growing several days prior to the weather event (at all noise levels) and then decreases drastically on the day of the event. For example, at the lowest noise level introduced into the synthetic data the day-to-day change is close to zero three days prior to the event, but increases to 0.95 two days before the event and again increases to 2.55 one day before the event. From the day before the event to the day of the event there is a drop of 7.8 in the rate of searches on average.

\subsection{A heuristic for the choice of $\alpha$ and $\epsilon$}
\label{section:heuristics}

We have seen in Section~\ref{sec:sims} that the choice of $\alpha$ and $\epsilon$ governs how conservative our method is. Although it is ultimately up to the analyst to chose and justify the parameters he uses in the analysis, we provide some heuristics to guide this choice. In this section, we will let $Y^{obs}_{i,t}$ be either the observed outcomes, or the first order differences which we denoted by $Y^*$ in the previous section. Consider $\Delta_{i,t}$ and $\partial \Delta_i,t$ as in Section~\ref{sec:method}, then let $\bar{\Delta}$ and $\overline{\partial \Delta}$ be the respective averages of their absolute values, $\tilde{\Delta}$ and $\widetilde{\partial \Delta}$ the respective maxima of their absolute values, and  $\mbox{se}(\Delta)$ and $\mbox{se}(\partial \Delta)$ the respective standard errors of their absolute values. We suggest choosing values of $\alpha$ and $\epsilon$ satisfying:
\begin{equation}
	\alpha_{min} = \frac{\tilde{\Delta}}{\bar{\Delta} + 3 \mbox{se}(\Delta)} \leq \alpha \leq \frac{\tilde{\Delta}}{\bar{\Delta} + \mbox{se}(\Delta)} = \alpha_{max}
\end{equation}
and 
\begin{equation}
	\epsilon_{min} = \frac{\widetilde{\partial \Delta}}{\overline{\partial \Delta} + 3 \mbox{se}(\partial \Delta)} \leq \epsilon \leq \frac{\widetilde{\partial \Delta}}{\overline{\partial \Delta} + \mbox{se}(\partial \Delta)} = \epsilon_{max}
\end{equation}

These heuristics are based on the interpretation of $\alpha$ and $\epsilon$ as measures of variation in the outcomes and the first differences of the outcomes. The choice of $\alpha_{min}$ is the ratio of the maximum absolute variation in outcomes to three standard deviations more than the mean while $\alpha_{max}$ is the ratio of the maximum variation in outcomes to one standard deviation more than the mean. The smaller $\alpha$ values are thus associated with how extreme the maximum is in comparison to the mean. Similarly for $\epsilon$, smaller values are associated with how extreme the maximum of the absolute value of first differences in outcomes is in comparison to the mean. The larger $\alpha$ and $\epsilon$ values have similar interpretation but with respect to the less extreme single standard deviation from the mean. The choice of one and three standard deviations is motivated by normal asymptotics.

\begin{figure}
	\centering
	\includegraphics[scale=0.68]{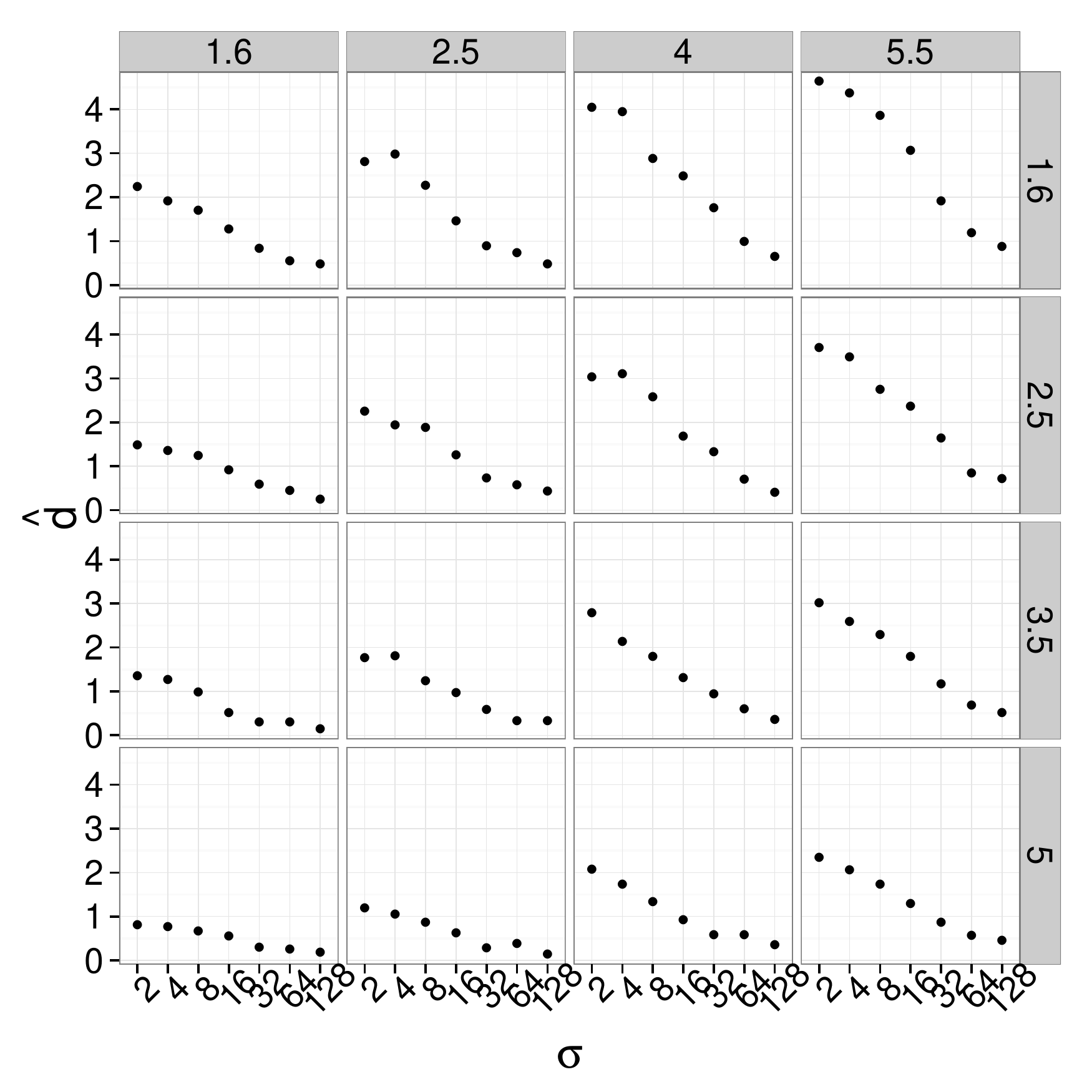}
	\caption{\onehalfspacing Values of $\hat{d}$ for a range of parameters $\alpha$ and $\epsilon$ obtained using the heuristics in Section~\ref{section:heuristics}, for different levels of noise}
	\label{fig:heuristics}
\end{figure}

With our synthetic data, the ranges of $[\alpha_{min}, \alpha_{max}]$ and $[\epsilon_{min}, \epsilon_{max}]$ depend on the noise level $\sigma$ --- for illustration purposes, we consider the ranges $\alpha \in [1.6, 5.5]$ and $\epsilon \in [1.6, 5]$ obtained by the unions of the ranges for the different values $\sigma = 2^i, \,\,\, i=1\ldots7$. Figure~\ref{fig:heuristics} displays the values of $\hat{d}$ that would be obtained for different combinations of $\alpha$ and $\epsilon$ within the range of our heuristics. Note that the parameters $\alpha$ and $\epsilon$ do not affect the estimate of the effect, only their causal interpretation.



\section{Concluding remarks}
\label{sec:disc}

The methodology we developed in this article is not
intended to supersede any of the traditional methodology
for dealing with observational studies, but rather to 
complement it. At a very high level, one can see our method
as a pre-processing step, which provide the analyst with
one level of protection against unsubstantiated causal claims.

We make two main contributions. First we provide a set of
assumptions and a method to determine a window before the
day of the observed event for which we can we can make
causal statement. Our second contribution is cleanly separate
the overall process into a pilot study, which is used to determine the 
window in which we can make causal statements, and the
causal analysis, which is carried on a disjoint subset of data. This
precaution insulates the causal analysis from any dependence 
on the observed outcomes used int he analysis. We have shown
in simulation studies that our method becomes increasingly 
conservative when the observed outcomes become volatile,
and that passed a certain level of noise, the method precludes
any causal statement beyond the date when the treatment
proxy is observed.

Our methodology extends the reach of causal inference to a 
specific type of observational studies, in which it is suspected
that the causal effect happens before the date in which a proxy
to the treatment is observed. The price paid for this extension is 
a reliance on extra assumptions, and a loss of efficiency since the
causal analysis is carried only on a subset of data. We also
emphasize the fact that the causal analysis carried being an 
observational studies, it suffers from the usual limitations.



 
\bibliographystyle{plainnat}
\addcontentsline{toc}{section}{References}
\bibliography{biblio}

\begin{thebibliography}{17}
\providecommand{\natexlab}[1]{#1}
\providecommand{\url}[1]{\texttt{#1}}
\expandafter\ifx\csname urlstyle\endcsname\relax
  \providecommand{\doi}[1]{doi: #1}\else
  \providecommand{\doi}{doi: \begingroup \urlstyle{rm}\Url}\fi

\bibitem[Angrist et~al.(2000)Angrist, Graddy, and Imbens]{angrist2000aa}
J.~D. Angrist, K~Graddy, and G.~W. Imbens.
\newblock The interpretation of instrumental variables estimators in
  simultaneous equations models with an application to the demand for fish.
\newblock \emph{Review of Economics Studies}, 67\penalty0 (3):\penalty0
  499--527, 2000.

\bibitem[Dunning(2012)]{dunning_natural_2012}
Thad Dunning.
\newblock \emph{Natural Experiments in the Social Sciences: A Design-Based
  Approach}.
\newblock Cambridge University Press, Cambridge ; New York, October 2012.
\newblock ISBN 9781107698000.

\bibitem[Greiner and Rubin(2011)]{greiner2011causal}
D~James Greiner and Donald~B Rubin.
\newblock Causal effects of perceived immutable characteristics.
\newblock \emph{Review of Economics and Statistics}, 93\penalty0 (3):\penalty0
  775--785, 2011.

\bibitem[Imbens and Rubin(2015)]{imbens2014causal}
Guido Imbens and Donald~B. Rubin.
\newblock \emph{Causal inference for statistics, social, and biomedical
  sciences : an introduction}.
\newblock Cambridge University Press, New York, 2015.
\newblock ISBN 978-0521885881.

\bibitem[Lewis et~al.(2011)Lewis, Rao, and Reiley]{lewis2011here}
Randall~A Lewis, Justin~M Rao, and David~H Reiley.
\newblock Here, there, and everywhere: correlated online behaviors can lead to
  overestimates of the effects of advertising.
\newblock In \emph{Proceedings of the 20th international conference on World
  wide web}, pages 157--166. ACM, 2011.

\bibitem[Murray et~al.(2010)Murray, Di~Muro, Finn, and
  Leszczyc]{murray2010effect}
Kyle~B Murray, Fabrizio Di~Muro, Adam Finn, and Peter~Popkowski Leszczyc.
\newblock The effect of weather on consumer spending.
\newblock \emph{Journal of Retailing and Consumer Services}, 17\penalty0
  (6):\penalty0 512--520, 2010.

\bibitem[Phan and Airoldi(2015)]{phan2015aa}
T.~Q. Phan and E.~M. Airoldi.
\newblock A natural experiment of social network formation and dynamics.
\newblock \emph{Proceedings of the National Academy of Sciences}, 112\penalty0
  (21):\penalty0 6595--6600, 2015.

\bibitem[Rosenbaum(2002)]{Rosenbaum:2002aa}
P.~R. Rosenbaum.
\newblock \emph{Observational Studies}.
\newblock Springer, 2nd edition, 2002.

\bibitem[Rosenbaum(2010)]{Rosenbaum:2010aa}
P.~R. Rosenbaum.
\newblock \emph{Design of Observational Studies}.
\newblock Springer, 2nd edition, 2010.

\bibitem[Rosenbaum and Rubin(1983)]{rosenbaum1983assessing}
Paul~R Rosenbaum and Donald~B Rubin.
\newblock Assessing sensitivity to an unobserved binary covariate in an
  observational study with binary outcome.
\newblock \emph{Journal of the Royal Statistical Society. Series B
  (Methodological)}, pages 212--218, 1983.

\bibitem[Rubin(1991)]{rubin1991aa}
D.~B. Rubin.
\newblock Practical implications of modes of statistical inference for causal
  effects and the critical role of the assignment mechanism.
\newblock \emph{Biometrics}, 47:\penalty0 1213--1234, 1991.

\bibitem[Rubin(1974)]{rubin1974estimating}
Donald~B Rubin.
\newblock Estimating causal effects of treatments in randomized and
  nonrandomized studies.
\newblock \emph{Journal of educational Psychology}, 66\penalty0 (5):\penalty0
  688, 1974.

\bibitem[Rubin(1996)]{rubin1996multiple}
Donald~B Rubin.
\newblock Multiple imputation after 18+ years.
\newblock \emph{Journal of the American statistical Association}, 91\penalty0
  (434):\penalty0 473--489, 1996.

\bibitem[Splawa-Neyman et~al.(1990)Splawa-Neyman, Dabrowska, Speed,
  et~al.]{splawa1990application}
Jerzy Splawa-Neyman, DM~Dabrowska, TP~Speed, et~al.
\newblock On the application of probability theory to agricultural experiments.
  essay on principles. section 9.
\newblock \emph{Statistical Science}, 5\penalty0 (4):\penalty0 465--472, 1990.

\bibitem[Starr-McCluer(2000)]{starr2000effects}
Martha Starr-McCluer.
\newblock \emph{The effects of weather on retail sales}.
\newblock Divisions of Research \& Statistics and Monetary Affairs, Federal
  Reserve Board, 2000.

\bibitem[Wager and Athey(2015)]{wager2015estimation}
Stefan Wager and Susan Athey.
\newblock Estimation and inference of heterogeneous treatment effects using
  random forests.
\newblock \emph{arXiv preprint arXiv:1510.04342}, 2015.

\bibitem[Zwebner et~al.(2013)Zwebner, Lee, and
  Goldenberg]{zwebner2013temperature}
Yonat Zwebner, Leonard Lee, and Jacob Goldenberg.
\newblock The temperature premium: Warm temperatures increase product
  valuation☆.
\newblock \emph{Journal of Consumer Psychology}, 24\penalty0 (2):\penalty0
  251--259, 2013.

\end{thebibliography}
\end{document}